\documentclass[12pt,preprint]{aastex}

\usepackage{natbib}
\usepackage{rotating}
\usepackage{bm}

\begin{document}

\title{On the Probable Existence of an Abrupt Magnetization in the Upper Chromosphere of the Quiet Sun}

\slugcomment{\bf Accepted for publication in The Astrophysical Journal Letters (2010).}

\author{Ji\v{r}\'{\i} \v{S}t\v{e}p\'an\altaffilmark{1,2}, Javier Trujillo Bueno\altaffilmark{1,3}}
\altaffiltext{1}{
Instituto de Astrof\'{\i}sica de Canarias, V\'\i a L\'actea s/n,
E-38205 La Laguna, Tenerife, Spain
}
\altaffiltext{2}{Associate Scientist at Astronomical Institute ASCR, v.v.i.,
Ond\v{r}ejov, Czech Republic}
\altaffiltext{3}{Consejo Superior de Investigaciones
Cient\'{\i}ficas (Spain)}
\email{stepan@iac.es; jtb@iac.es}

\begin{abstract}
We report on a detailed radiative transfer modeling of the observed scattering polarization in the H${\alpha}$ line, which allows us to infer quantitative information on the magnetization of the quiet solar chromosphere. Our analysis suggests the presence of a magnetic complexity zone with a mean field strength $\langle B \rangle \, {>} \, 30$\,G lying just below the sudden transition region to the coronal temperatures. The chromospheric plasma directly underneath is very weakly magnetized, with $\langle B \rangle \, {\sim} \, 1$\,G. The possible existence of 
this abrupt change in the degree of magnetization of the upper chromosphere of the quiet Sun might have large significance for our understanding of chromospheric (and, therefore, coronal) heating. 
\end{abstract}


\section{Introduction\label{sec:intro}}

A long-standing issue in solar astrophysics concerns the strength and structure variations with height of the magnetic field in the chromosphere of the quiet Sun \citep[e.g.,][]{judge06,harvey09}. Our empirical knowledge on this issue has remained vague notwithstanding the qualitative information provided by high resolution monochromatic images of the solar atmosphere taken at various wavelengths across strong spectral lines like H${\alpha}$. For example, H${\alpha}$ line-center images show a mass of cell-spanning fibrils as a flattened carpet, with upright ones jutting out from network patches \citep{rutten07}. Unfortunately, such images at various wavelengths ($\lambda$) across the {\em intensity} profiles ($I(\lambda)$) of chromospheric lines do not provide quantitative information on the magnetic field vector.

The temperature minimum region of solar atmospheric models is transparent to H${\alpha}$ radiation \citep{schoolman72}. As a result, we see the chromosphere at the very line center of H${\alpha}$ but the photosphere in the line wings. It is thus not surprising that the response function of the emergent circular polarization induced by the Zeeman effect to magnetic field perturbations shows large photospheric contributions \citep{socas04}. On the contrary, in the quiet Sun the linear polarization observed in H${\alpha}$ and in many other spectral lines is fully dominated by the presence of radiatively induced population imbalances and quantum coherences among the magnetic sublevels of the line's levels \citep[e.g.,][]{jtb09rev}, which produce $Q/I$ profiles whose maximum values are usually located at the line center \citep{stenflo97,gandorfer00}. Moreover, this scattering line polarization is modified by the Hanle effect, which operates mainly in the line core and gives rise to $Q/I$ and $U/I$ profiles different to those corresponding to the zero-field case \citep[e.g.,][]{stenflobook,ll04}. Therefore, the Hanle effect in strong lines like H${\alpha}$ is the physical mechanism that should be exploited for facilitating quantitative explorations of the magnetism of the ``quiet" solar chromosphere.

\begin{figure}
\plotone{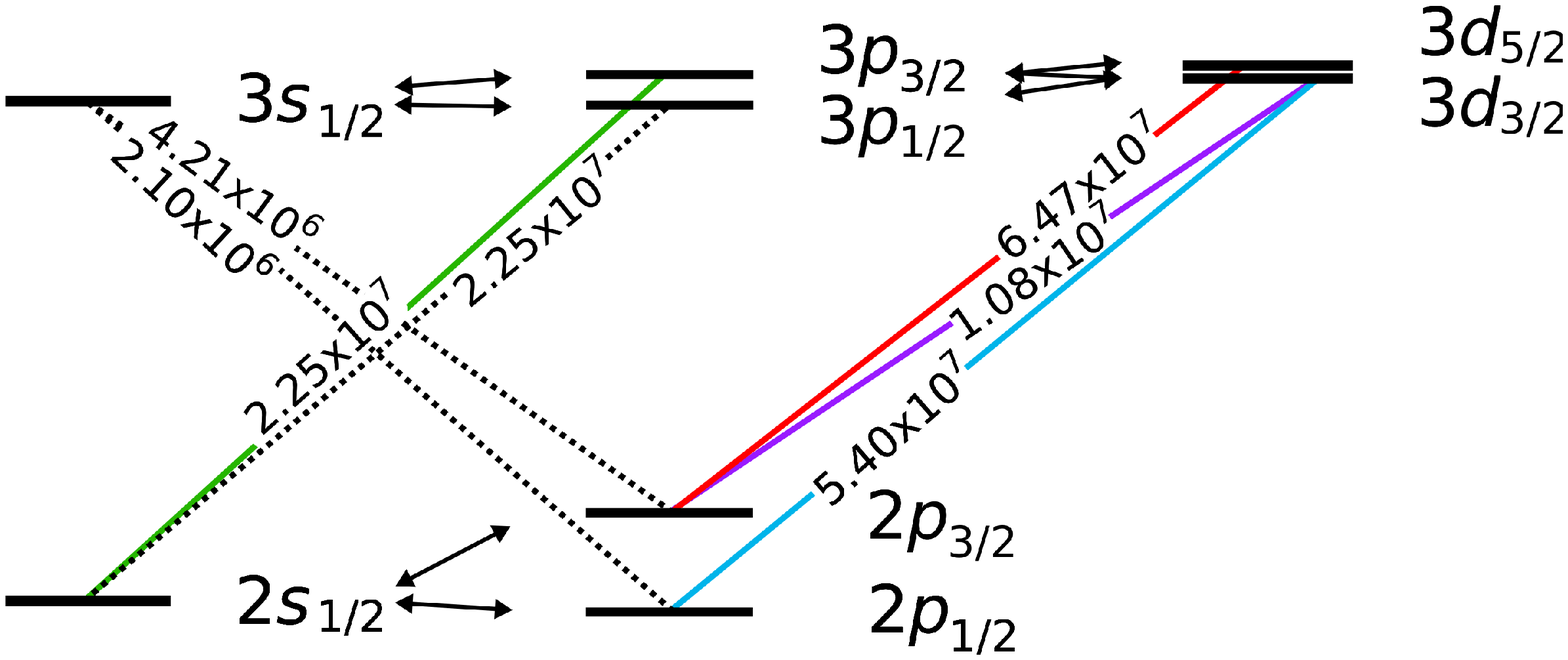}
\caption{
Components of the H$\alpha$ line. Only 4 ({\it solid lines}) out of 7 ({\it solid and dotted lines}) blended transitions make a meaningful contribution to the scattering polarization of the H${\alpha}$ line in the solar atmosphere. The atomic polarization of the $2p_{3/2}$ level turns out to be negligible in the H$\alpha$ formation region; hence selective absorption of polarization components \citep[e.g.,][]{manso-trujillo-03} 
does not contribute to the linear polarization of H$\alpha$ in the quiet Sun. The Einstein $A_{ul}$ coefficients are given in s$^{-1}$. The arrows indicate the collisional dipolar depolarizing transitions. The three upper levels with angular momentum $j>1/2$ have the following critical Hanle fields: $B_H(3p_{3/2})$=16\,G, $B_H(3d_{3/2})$=9\,G and $B_H(3d_{5/2})$=6\,G, where $B_H$ is the magnetic strength for which the level's Zeeman splitting equals its natural width. 
}
\label{fig:grotr}
\end{figure}

In the quiet Sun the fractional linear polarization signals of the H${\alpha}$ line are weak (i.e., of the order of $0.1\%$ when observing close to the solar limb), so that their detection with the present telescopes requires sacrifice the spatio-temporal resolution to be able to achieve high polarimetric sensitivity. As a result, it is easier to detect Stokes $Q$ (with the tangent to the nearest solar limb being its reference direction) than Stokes $U$ signals (whose positive and negative values quantify the rotation of the direction of linear polarization). For these reasons, typically only the Stokes $Q/I$ profile is available, as is indeed the case with the H${\alpha}$ observation by \citet{gandorfer00} that will be interpreted here. One could thus think that without $U/I$ it is presently impossible to infer the presence of an unresolved magnetic field in the quiet solar chromosphere, because it would require confronting the observed $Q/I$ amplitude in the H${\alpha}$ line with the one that the highly inhomogeneous and dynamic solar chromospheric plasma  would produce if it were unmagnetized. This strategy could be applied successfully for determining the mean magnetization of the quiet solar photosphere by solving the radiative transfer problem for the Sr\,{\sc i} at 4607\,\AA\ line in a realistic three-dimensional (3D) hydrodynamical model of the quiet solar photosphere \citep{jtbNat04}, but a similar approach for inferring instead the magnetization of the quiet chromospheric plasma is not yet possible mainly because it is still computationally prohibitive to produce a realistic 3D model of the thermal, density and dynamic structure of the quiet chromosphere \citep[e.g.,][]{leenaarts10}. Fortunately, the $Q/I$ profile observed by \citet{gandorfer00} in a quiet region at about 5$''$ from the solar limb (i.e., at $\mu=\cos\theta\,{\approx}\,0.1$, with $\theta$ being the heliocentric angle) shows a peculiar line core asymmetry (LCA; see the wiggly line in any of the right panels of Fig.\,\ref{fig:fit}) whose very existence cannot be explained by the mere fact that the H${\alpha}$ line results from the superposition of seven components, four of which contribute to Stokes $Q$ (see Fig.\,\ref{fig:grotr}). Moreover, the $Q/I$ profile of the (photo-ionization dominated) H${\alpha}$ line is not very sensitive to the chromospheric thermal structure, which justifies the strategy adopted here of using various (hot and cool) one-dimensional semi-empirical models of the quiet chromosphere for investigating the physical origin of the observed LCA. It is also necessary to note that the observed intensity profile \citep[see][]{gandorfer00}, which is fully insensitive to the magnetic field of the quiet solar chromosphere, does not show any noteworthy line-core asymmetry and that the small atomic weight of hydrogen does not favor broadening by macroscopic fluid motions. As shown in this letter, the LCA present in the observed $Q/I$ profile can be explained by the Hanle effect of a magnetic field in the solar atmosphere whose height variation suggests the presence of a magnetic complexity zone located in the upper chromosphere of the quiet Sun.

\section{The Radiative Transfer Problem}

In the absence of collisional depolarization the critical value of the magnetic strength (in gauss) for the onset of the Hanle effect in a given atomic level is $B_H{=}1.137{\times}10^{-7}/(t_{\rm life}g)$, with $g$ the level's Land\'e factor and $t_{\rm life}$ its radiative lifetime in seconds (see the legend of Fig.\,\ref{fig:grotr}). Typically, the Hanle effect is sensitive to magnetic strengths between $0.1B_H$ and $10B_H$, but note that due to collisional quenching the real $B_H$ values of the $nlj$ levels of hydrogen can be significantly larger. In fact, collisions with protons and electrons are efficient in producing transitions between the ($n,l,j$) and the ($n,l{\pm}1,j^{'}$) levels. Such collisional transitions play a non-negligible depolarizing role because at some heights their rates are similar to the inverse radiative lifetimes of the hydrogen levels. For $B>10\,B_H(j_u)$ (with $B_H(j_u)$ being the $B_H$ value of the line's upper level) the Hanle effect is only sensitive to the orientation of the magnetic field vector, but not to its strength (the saturation regime of the Hanle effect).

Since the quiet solar chromosphere is optically thick at the center of the H${\alpha}$ line, we need to take into account radiative transfer (RT) effects. To compute the emergent $Q/I$ profile in models of the quiet solar atmosphere we have applied a computer program \citep{thesis08} that takes into account the overlapping of the hydrogen transitions when solving jointly the statistical equilibrium equations for the multipolar components of the atomic density matrix and the Stokes-vector transfer equations \citep[see Chapter 7 in][]{ll04} via a fast iterative method and accurate formal solver \citep{jtb03}. As justified in \S1, we have used various (hot and cool) one-dimensional semi-empirical models of the quiet solar atmosphere assuming that at each height, $z$, there is a random-azimuth magnetic field of strength $B(z)$, inclined by an angle $\theta_B(z)$ with respect to the local vertical. Our conclusions on the physical origin of the observed LCA 
are based on multilevel RT calculations in such models,  
and it suffices with showing in the following section 
the results for the well-known FAL-C semi-empirical model \citep{fontenla93}. The chosen atomic model includes all the fine structure levels corresponding to the first four $n$-levels of the hydrogen atom. The neglect of hyperfine structure (HFS) should not produce any sizable overestimation of the scattering polarization amplitudes \citep{bommier82}. We quantify the excitation state of each $nlj$ level by means of the multipolar components of its corresponding atomic density matrix \citep{ll04}. Such $\rho^K_Q(nlj)$ elements (with $0\le K\le 2j$ and $-K\le Q\le K$) quantify the level's overall population ($\rho^0_0$), the population imbalances among its sublevels ($\rho^{K>0}_0$) and the quantum coherences between each pair of them ($\rho^{K>0}_{Q\neq 0}$). We neglect quantum coherences between different $j$ levels. This is a good approximation for investigating the Hanle effect in the H${\alpha}$ line, assuming that the Hanle effect is produced by the only action of magnetic fields of strength $B{\lesssim}100$\,G. The reason is that under such circumstances the selection rule $\Delta{l}={\pm}1$ for radiative transitions inhibits radiative couplings between levels ($n,l,j$) and ($n,l{\pm}1,j^{'}$). Moreover, for $B{\lesssim}100$\,G the energy separation between levels ($n,l,j$) and ($n,l{\pm}2,j^{'}$) is larger than both their natural width and than the ensuing Zeeman splitting. 

As mentioned above, the consideration of isotropic collisions with protons and electrons is important for a rigorous modeling of the scattering polarization in hydrogen lines. We have applied the semi-classical impact approximation theory \citep{sahal96} for calculating the dominant dipolar collisional rates between the fine-structure levels. The Stark broadening of the hydrogen lines is accounted for by using a suitable approach for the hydrogen lines \citep{stehle96}. The inelastic collisional rates between the different $n$-levels were also taken into account \citep{przybilla04}. With these physical ingredients our non-LTE synthesis of the H${\alpha}$ intensity profile is in good agreement with that obtained by other researchers \citep[e.g.,][]{przybilla04b} using an atomic model with more $n$-levels.

Finally, it is also important to clarify that our calculations are based on the complete frequency redistribution (CRD) theory of spectral line polarization \citep{ll04} which might be considered unsuitable for the H${\alpha}$ line given that the electron and proton densities are sensitive to partial redistribution effects through multilevel interlockings with the other transitions included in our atomic model \citep{hubeny95}. However, in our calculations we have used the electron and proton densities of the FAL-C model itself and the Lyman lines tend to be optically thick in the atmospheric region of the H${\alpha}$ line formation.  Under such circumstances the CRD approximation is a suitable one for subordinate lines like H${\alpha}$ \citep[e.g.,][]{heinzel95}.

\section{The Physical Interpretation\label{sec:interp}}

The emergent fractional linear polarization $Q(\lambda)/I(\lambda)$ profile of the H$\alpha$ line is basically the result of the superposition of four $Q_{ij}(\lambda)/I(\lambda)$ blended components, where $I(\lambda)$ is the total intensity profile and each $Q_{ij}(\lambda)$ profile can be obtained by solving the transfer equation for Stokes $Q$ using the $I$-component of the total absorption coefficient ($\eta_I$) and the component's emissivity $\epsilon^{ij}_Q$ (which is symmetric with respect to the component's central wavelength): $Q_{ij}(\lambda){\approx}\int_0^\infty (\epsilon^{ij}_Q/\eta_I) {\rm e}^{-t}\,{\rm d} t$, with $t$ the total optical thickness along the line of sight (LOS). As the fine-structure components $i$--$j$ of the H${\alpha}$ line are shifted from line center (see Fig.\,\ref{fig:grotr}), the optical depth scale is not symmetric with respect to the component central wavelength $\lambda_{ij}$ (i.e., $\eta_I(\lambda_{ij}-\Delta\lambda)\neq\eta_I(\lambda_{ij}+\Delta\lambda)$). It follows that, in general, $Q_{ij}(\lambda_{ij}-\Delta\lambda)\,{\ne}\,Q_{ij}(\lambda_{ij}+\Delta\lambda)$ so that the individual linear polarization components of the H$\alpha$ line are asymmetric with respect to $\lambda_{ij}$. In conclusion: (1) The $Q_{ij}/I$ profiles are asymmetric even in the absence of magnetic field; (2) the formation depth of the $Q_{ij}/I$ components is a non-trivial function of wavelength. 

\begin{figure*}
\plottwo{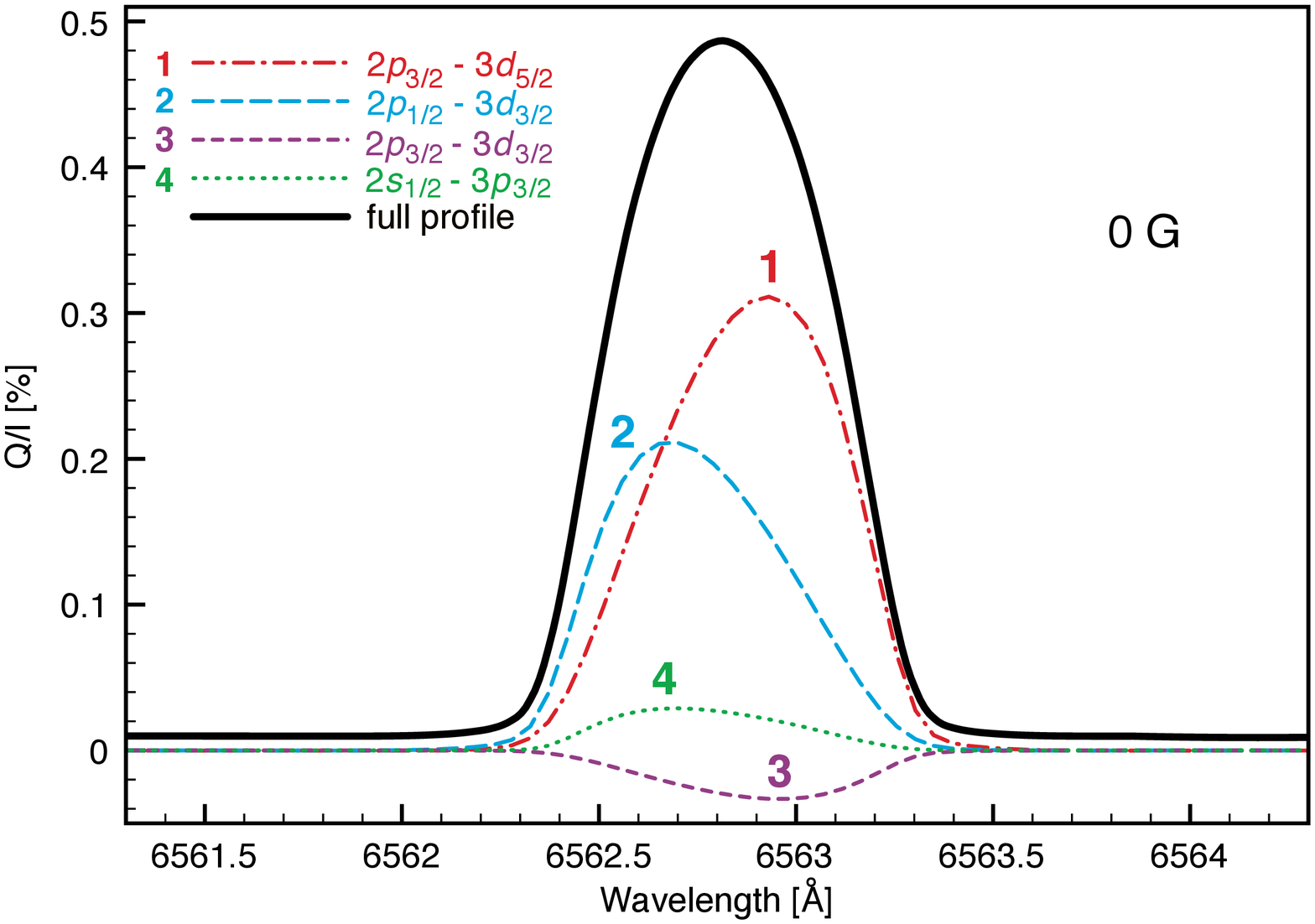}{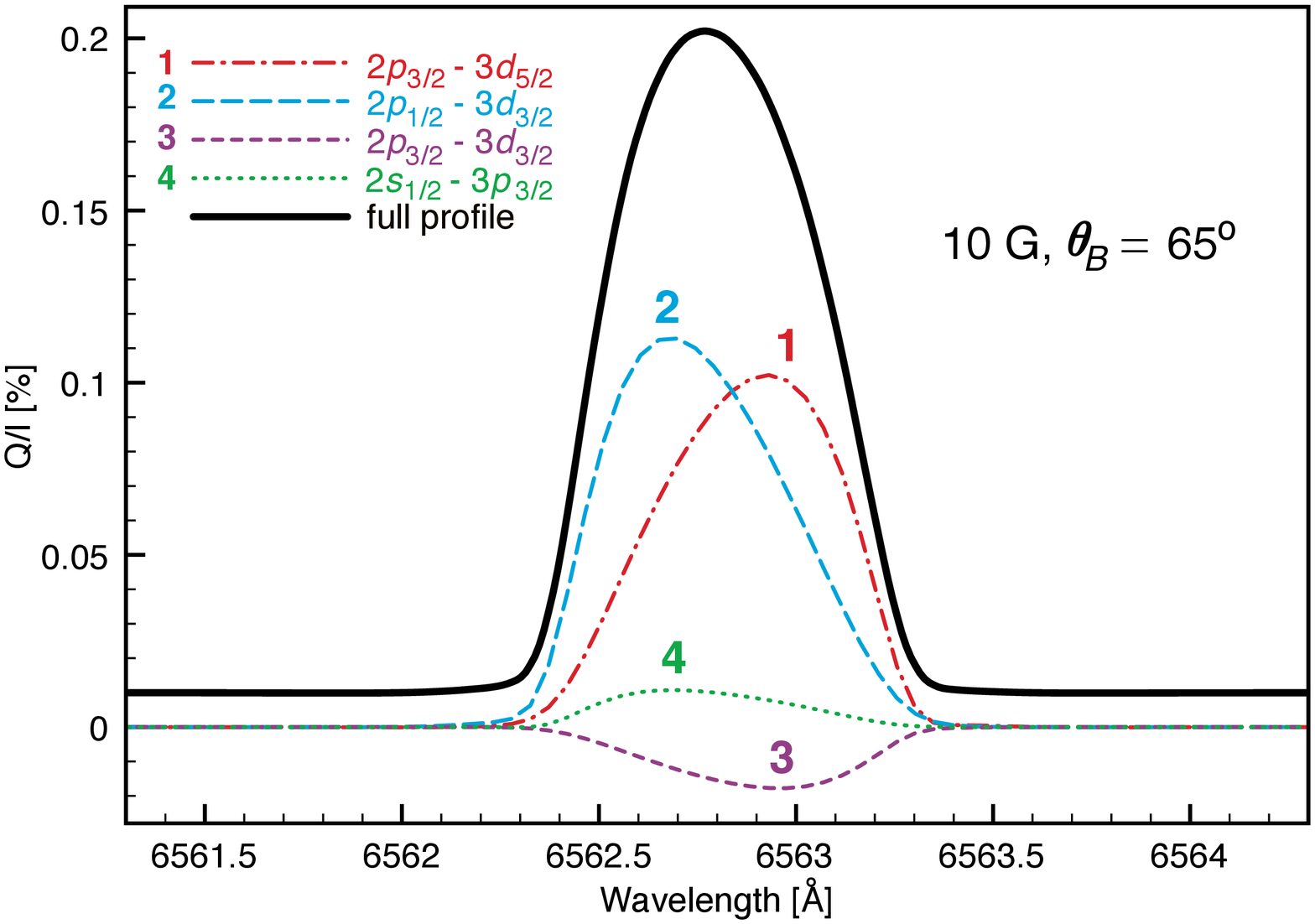}
\caption{
$Q/I$ profiles of the H$\alpha$ line and of its four polarizing components calculated in the FAL-C model atmosphere for a LOS with $\mu=0.1$. In each panel the full $Q/I$ profile includes the contribution of the continuum polarization. The positive reference direction for Stokes $Q$ is the parallel to the nearest limb. {\it Left panel}: the $B=0$\,G case. We point out that identical $Q/I$ shapes are obtained in the presence of a random-azimuth magnetic field with a strength that produces saturation of the upper-level Hanle effect (e.g., a horizontal field with $B{>}50$\,G). The only difference is that the amplitude of the emergent polarization is smaller by a factor $(3\cos^2{\theta_B}-1)^2/4$ (e.g., it is $0.12\%$ for a horizontal magnetic field). {\it Right panel}: the emergent $Q/I$ profiles in the presence of a random-azimuth magnetic field with constant strength $B=10$\,G and inclination $\theta_B=65^{\circ}$. Note that in this case the ``2'' and ``3'' transitions play now a more important role, in addition to transition ``1''. In particular, transition ``2'' dominates the blue part of the full $Q/I$ profile while transition ``3'' decreases the profile on the red side. As a result, the whole $Q/I$ profile is slightly skewed toward blue wavelengths. However, no LCA similar to the observed one is produced because all components are depolarized in a rather wavelength-independent manner. 
}
\label{fig:unif}
\end{figure*}

Figure\,\ref{fig:unif} shows that within the framework of our radiative transfer modeling 
the LCA in the observed $Q/I$ profile cannot be explained neither in the absence of magnetic fields nor in the presence of homogeneous magnetic fields. The reason is that the shapes of the four $Q_{ij}/I$ components are smooth ``skewed gaussians'' whose superposition gives rise to a rather smooth $Q/I$ profile without any sizable asymmetry. The above-mentioned second conclusion provides the clue for our explanation of the observed LCA. In agreement with our numerical experiments, the LCA can be created by a modification of the $Q_{ij}/I$ profiles of some components (mainly the ``1'' and ``2'' transitions of Fig.\,\ref{fig:unif}) induced by a spatially varying magnetic field. The superposition of the blended linear polarization profiles can then give rise to a sizable LCA whose shape depends sensitively on the spatial distribution of the chromospheric magnetic field. 

Given that magnetic field models with a strength that decreases with height are also unable to explain the observed LCA we start with the following height-variation of the magnetic strength: 
\begin{equation}
B(z)=\frac{B_0}{1+{\rm e}^{-(z-z_0)/w}}\,+\,B_{b}\,.
\label{eq:sigmoid}
\end{equation}
This function, whose first term is a sigmoid, tends to $B_0+B_b$ for $z\to+\infty$ (upper chromosphere) and to $B_b$ for $z\to{-}\infty$ (lower photosphere). Note that $B(z_0)=B_b+B_0/2$ and that the smaller the positive parameter $w$ the steeper the slope of the sigmoid. 

\begin{figure*}
\plottwo{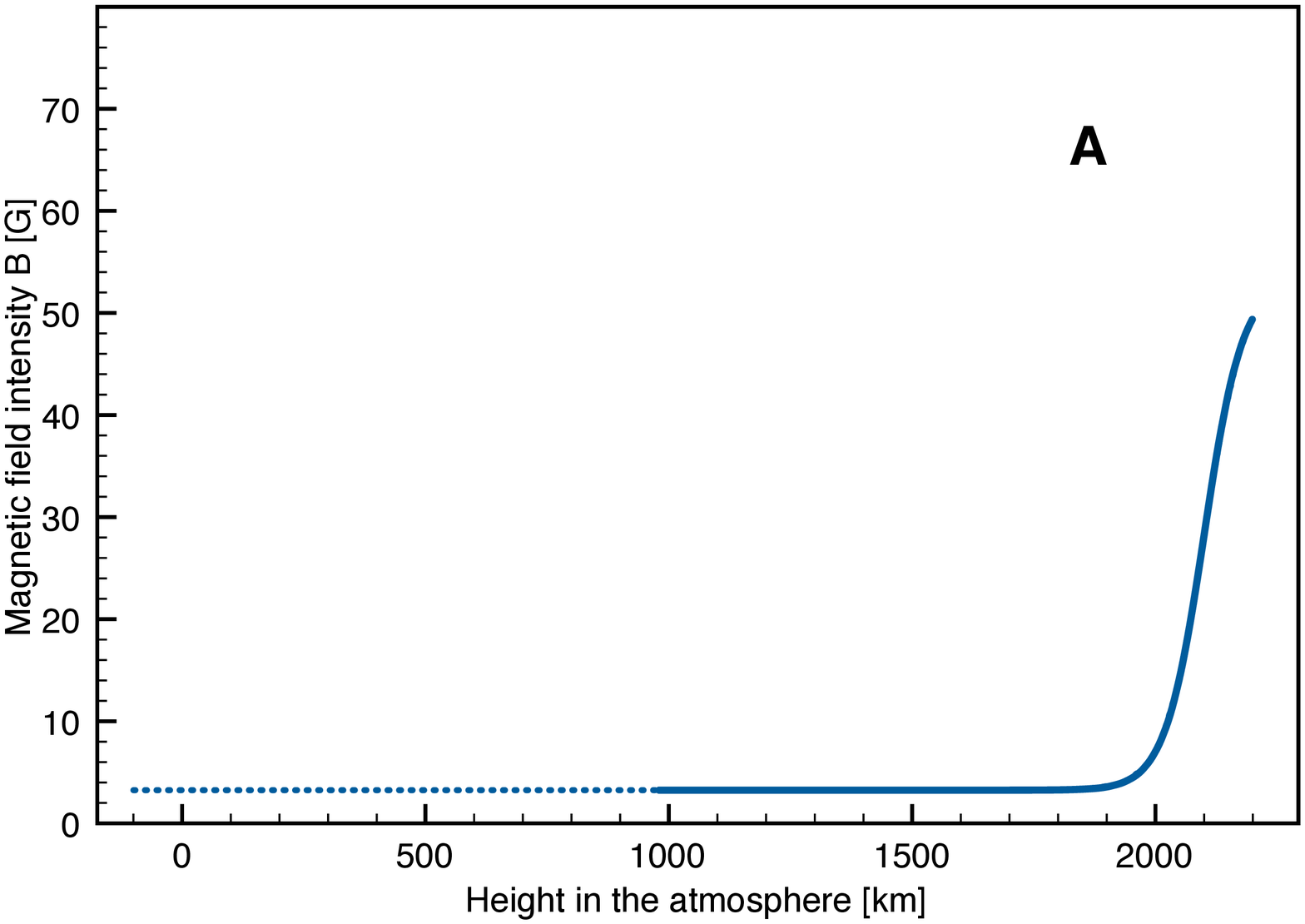}{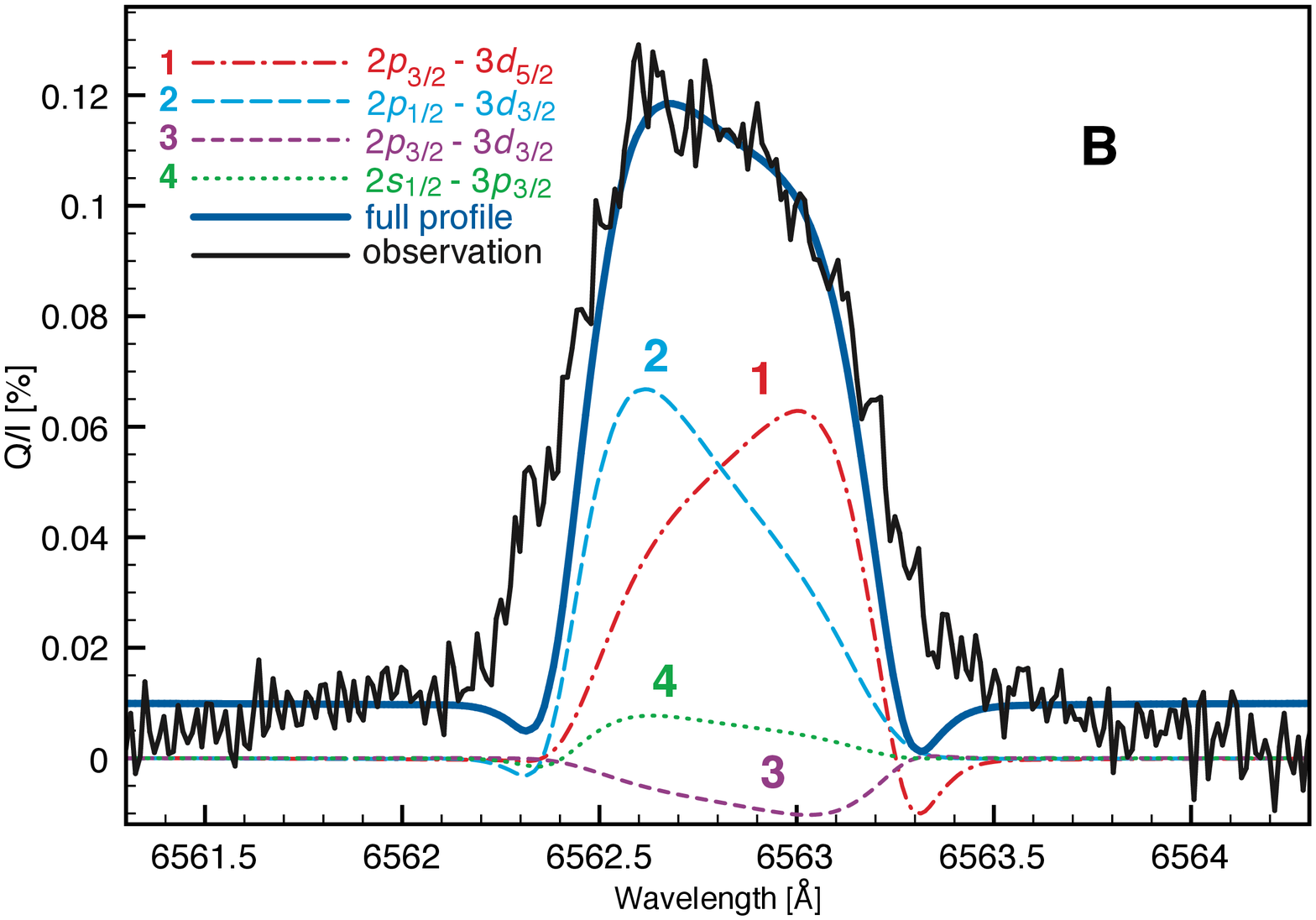}
\plottwo{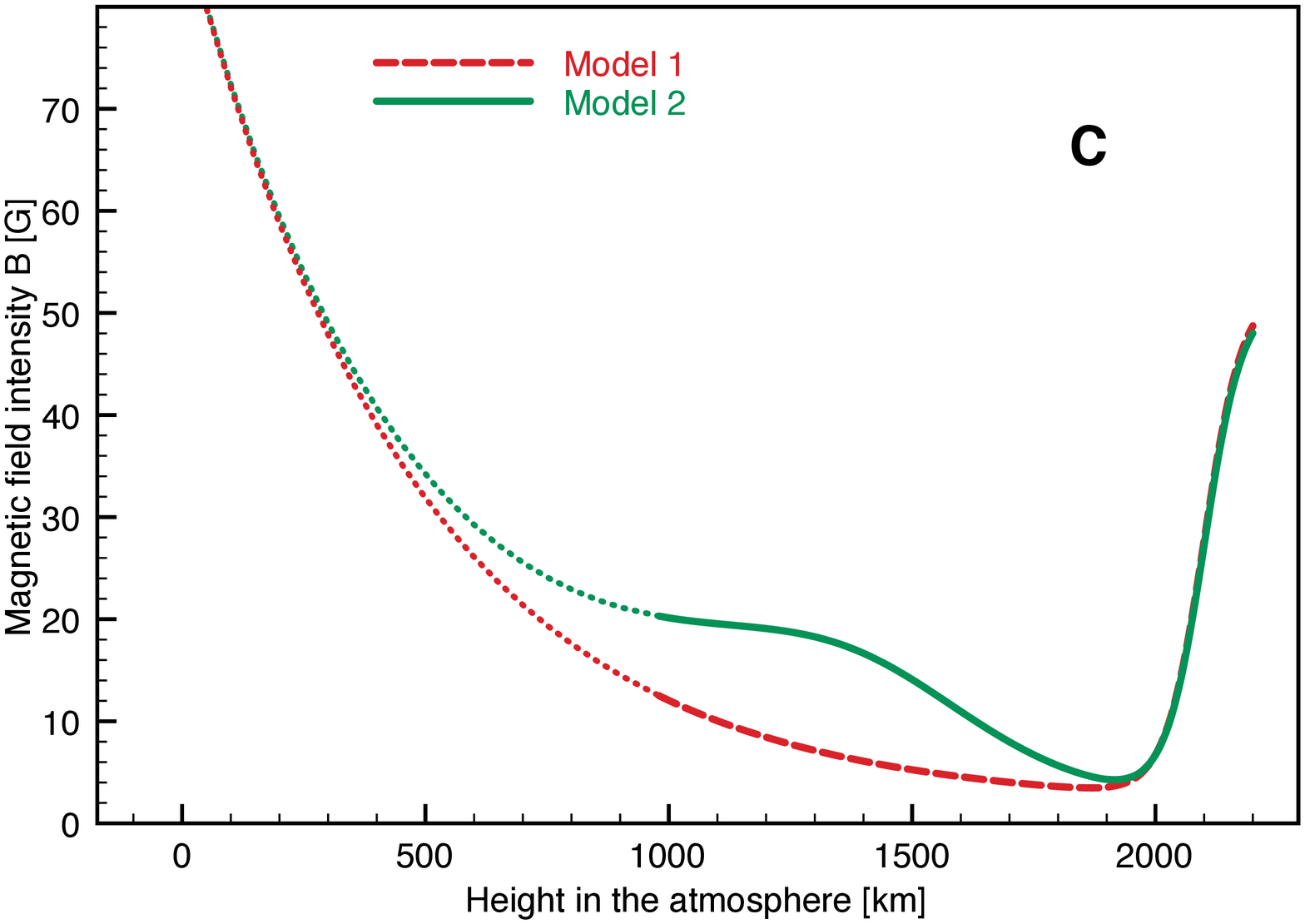}{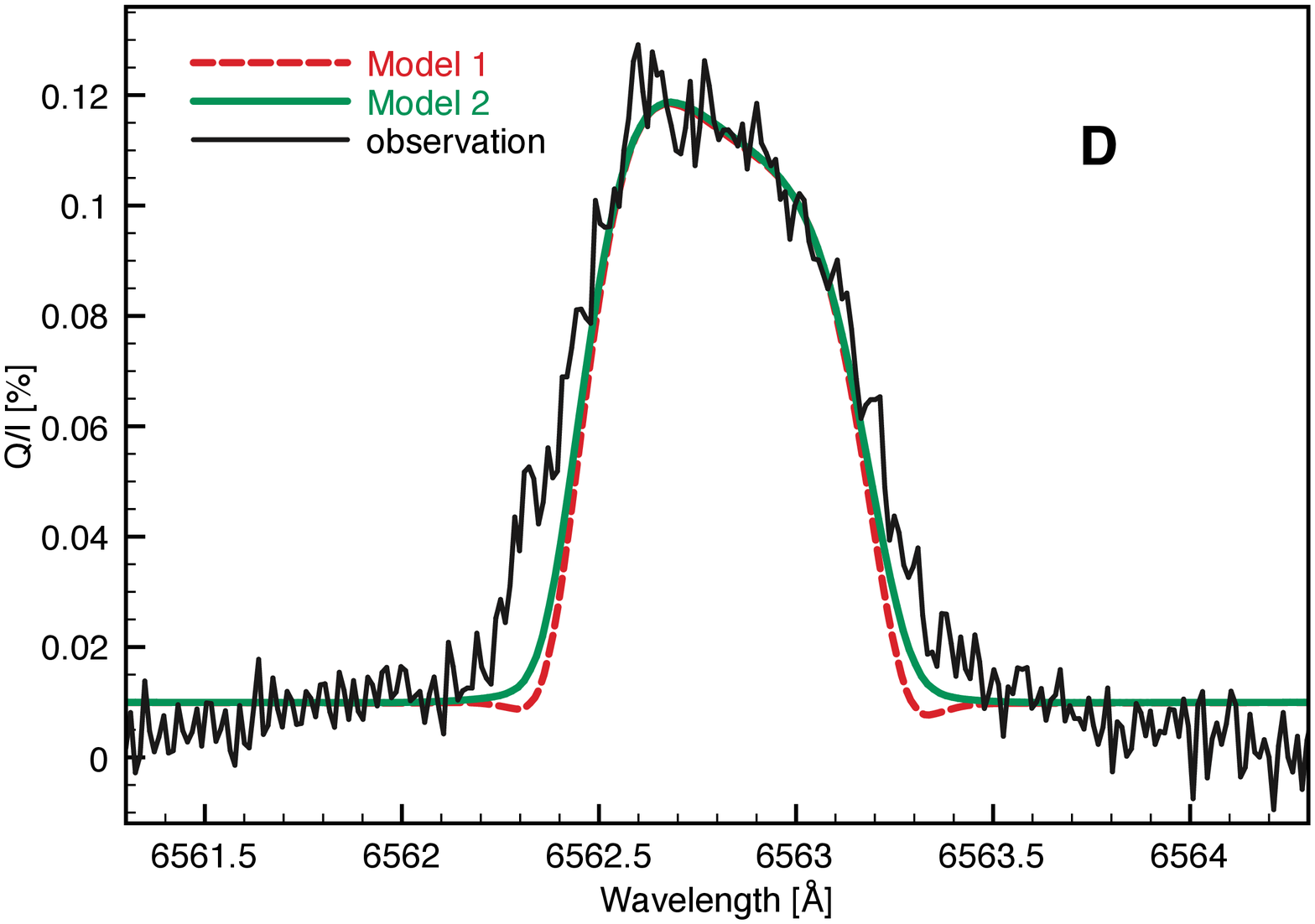}
\caption{
Magnetic field models ({\it left panels}) and calculated vs. observed $Q/I$ profiles ({\it right panels}). The wiggly line in the right panels shows the $Q/I$ profile of the H${\alpha}$ line observed by Gandorfer (2000) in a quiet region at about 5$''$ from the solar limb. Panel {\it B} shows the calculated $Q/I$ profile corresponding to the magnetic field model of panel {\it A}, while panel {\it D} shows the calculated $Q/I$ profiles corresponding to the magnetic field models of panel {\it C}. All these magnetic field models produce a LCA similar to that present in the observed $Q/I$ profile. Note in the left panels that for a LOS with $\mu=0.1$ the scattering polarization of the H${\alpha}$ line is insensitive to the structure of the magnetic field in the lower atmospheric region indicated by the dotted line part of the models. 
}
\label{fig:fit}
\end{figure*}

We fix for simplicity the inclination of our random-azimuth magnetic field and 
set $B_0=50$\,G because in order to model the observed $Q/I$ amplitude we need to achieve for the H$\alpha$ line practically full saturation of the Hanle effect in the upper chromosphere. From non-LTE model calculations in various semi-empirical models we have found that $z_0$ and $w$ have to be adjusted such that the field intensity in the upper chromosphere increases abruptly with height. For instance, in the FAL-C model the magnetic strength must increase from 3\,G at 1900\,km to 50\,G at 2200\,km, approximately. Note that in this semi-empirical model the transition region to the coronal temperatures starts at a height $z\,{\approx}\,2200$\,km and that magnetic fields at larger heights in this model do not affect the linear polarization of the H$\alpha$ emergent radiation for a LOS with $\mu=0.1$ because this line is already optically thin there. The other parameters of the magnetic field model are $\theta_B$ and $B_b$. Even though a horizontal field with a strength that increases with height in the upper chromosphere can induce a LCA, it does not seem to produce an optimal fit. We found a good agreement with the observed profile for $\theta_B=65^\circ$, $z_0=2100$\,km, $w=40$\,km and $B_b\,{\approx}\,3$\,G (see Fig.\,\ref{fig:fit}A). A comparison of the resulting $Q/I$ profile with the observed one can be found in Fig.\,\ref{fig:fit}B.

Although the agreement between the calculated and observed $Q/I$ profiles is very good around the line center, there is now disagreement in the line wings, which are formed deeper in the atmosphere. \citet{jtbNat04} showed that the bulk of the quiet solar photosphere is teeming with a distribution of tangled magnetic fields having a mean field strength ${\langle B \rangle}\,{\approx}\,60\,$\,G (when estimating ${\langle B \rangle}$ assuming the simplest case, adopted here, of a single value field strength). We have included this important physical ingredient by adding an extra sigmoid function in Eq.~(\ref{eq:sigmoid}) with a {\em positive} sign in the exponent, in order to get the $B(z)$ variation given by Model~1 of Fig.\,\ref{fig:fit}C. This model is still characterized by a magnetic zone with $\langle B \rangle \, {>} \, 30$\,G in the upper solar chromosphere, but has a strongly magnetized photosphere and a weakly magnetized lower chromosphere. The ensuing $Q/I$ profile is shown in Fig.\,\ref{fig:fit}D. As can be seen, the consideration of this additional ingredient to satisfy the constraint of a ``hidden"  magnetic field in the quiet solar photosphere leads to a slightly better overall fit to the fractional linear polarization observed in the H$\alpha$ line. We point out that this model for the height variation of the magnetic strength in the quiet solar atmosphere is not the only one that leads to a very good fit to the LCA of the observed $Q/I$ profile. The key point is however that all plausible models are qualitatively similar, in the sense that all of them are characterized by the presence of a significantly magnetized zone in the upper solar chromosphere (where $\langle B \rangle$ must be {\em at least} 30\,G), overlying weakly magnetized regions with $\langle B \rangle \, {\sim} \, 1$\,G. Model~2 of Fig.\,\ref{fig:fit}C shows a qualitatively similar magnetic field variation that leads almost to the largest possible further improvement in the H$\alpha$ line wings without deteriorating the LCA fit (see the ensuing $Q/I$ profile in Fig.\,\ref{fig:fit}D).

\section{Concluding Comments\label{sec:concl}}

The amplitude and shape of the scattering polarization profile of the H$\alpha$ line is very sentitive to the strength and structure of the magnetic field in the upper solar chromosphere. For this reason, we think that full Stokes-vector observations of the H$\alpha$ line within and outside coronal holes would probably lead to an important breakthrough in our empirical understanding of chromospheric magnetism. A suitable instrument with which we are carrying out such observations (in close collaboration with Dr. M. Bianda and Dr. R. Ramelli, from IRSOL) is the Z\"urich Imaging Polarimeter (ZIMPOL) attached to the Gregory Coud\'e Telescope of the Istituto Ricerche Solari Locarno (IRSOL) and to the telescope THEMIS of the Observatorio del Teide (Tenerife; Spain).

Our one-dimensional radiative transfer interpretation of the LCA in the $Q/I$ profile observed by \citet{gandorfer00} suggests the presence of an abrupt magnetization in the upper chromosphere of the quiet Sun. In all the semi-empirical models of the quiet Sun atmosphere we have considered, the base of this magnetic complexity zone needed to explain the observed LCA lies a few hundred kilometers below the height where the model's sudden transition region to the coronal temperatures is located. Although the solar chromosphere is highly inhomogeneous and dynamic, we believe that our suggestion should be taken seriously because H${\alpha}$ is a photoionization-dominated line whose observed Stokes $I(\lambda)$ profile (which is fully insensitive to the chromospheric magnetic field) does not show any noteworthy asymmetry within the line core interval where the observed $Q/I$ profile shows the peculiar asymmetry that motivated this investigation.

The probable existence of an abrupt change in the mean magnetization of the upper chromosphere of the quiet Sun may have large significance for the passage of emerged magnetized plasma across the chromosphere and into the corona \citep[e.g., the review by][]{morenoinsertis07} and for the overall energy balance of the outer solar atmosphere. For instance, coronal reconnection and dissipation of magnetic energy may be significantly modulated by the physical conditions in the upper chromosphere and by the height at which neighboring magnetic fibrils begin to press against each other \citep{parker07}. The presence of a strong and abrupt magnetization in the upper chromosphere of the quiet Sun might also hold the clue for clarifying the physical origin of the elusive acceleration mechanism of the solar wind.

\acknowledgments

We are grateful to R. Casini (HAO) for scientific discussions on the physics of the H${\alpha}$ polarization, which were very helpful to clarify that the asymmetry of the observed $Q/I$ profile is not caused by hyperfine structure effects. Thanks are also due to E. Landi Degl'Innocenti (University of Florence), R. Manso Sainz (IAC), F. Moreno Insertis (IAC) and S. Sahal-Br\'echot (Observatoire de Paris) for useful discussions. Financial support by the Spanish Ministry of Science (project AYA2007-63881) and by the SOLAIRE network (MTRN-CT-2006-035484) are gratefully acknowledged.

\end{document}